 \documentclass[final,5p,times,twocolumn]{elsarticle}

\usepackage{amssymb}
\usepackage{hyperref}
\hypersetup{colorlinks=true,%
           linkcolor=black,%
           anchorcolor=black,%
           citecolor=black,%
           filecolor=black,%
           menucolor=black,%
           urlcolor=black}

\usepackage{lineno}

\usepackage{subfigure}

\usepackage{stfloats}
\usepackage{float}

\newcommand{\neqcm}{\ensuremath{\mathrm{n}_{\mathrm{eq}}/\mathrm{cm}^2}}
\newcommand{\mum}{\ensuremath{\mu}m}

\journal{Nuclear Instruments and Methods A }
\begin{document}

\begin{frontmatter}


\title{Thin n-in-p pixel sensors and the SLID-ICV vertical integration 
technology for the ATLAS upgrade at the HL-LHC}
 
\author{A.~Macchiolo\corref{cor1}\fnref{label1}}
\ead{Anna.Macchiolo@mpp.mpg.de}
\author{L.~Andricek \fnref{label1,label2}}
\author{M.~Ellenburg\fnref{label1}}
\author{H.G.~Moser \fnref{label1,label2}}
\author{R.~Nisius\fnref{label1}}
\author{R.H.~Richter \fnref{label1,label2}}
\author{S.~Terzo\fnref{label1}}
\author{P.~Weigell\fnref{label1}}
\address[label1]{Max-Planck-Institut f\"ur Physik, F\"ohringer Ring 6, D-80805 M\"unchen, Germany}
\address[label2]{Max-Planck-Institut Halbleiterlabor, Otto Hahn Ring 6, D-81739 M\"unchen, Germany}


\author{}

\address{}

\begin{abstract}
The R\&D activity presented is focused on the development of new modules for the upgrade of the ATLAS pixel system at the High Luminosity LHC (HL-LHC). The performance after irradiation of n-in-p pixel sensors of different active thicknesses is studied, 
together with an investigation of a novel interconnection technique offered by the Fraunhofer Institute EMFT in Munich, 
the Solid-Liquid-InterDiffusion (SLID), which is an alternative to the standard solder bump-bonding. 
The pixel modules are based on thin n-in-p sensors, 
with an active thickness of 75 \mum\, or 150 \mum, 
produced at the MPI Semiconductor Laboratory (MPI HLL) and on 100 \mum \, thick sensors with active edges, fabricated at VTT, Finland. 
Hit efficiencies are derived from beam test data for thin devices irradiated up to a fluence of  $4\cdot 10^{15} \neqcm$.
For the active edge devices, the charge collection properties of the edge pixels before irradiation
is discussed in detail, with respect to the inner ones, using measurements with radioactive sources. 
Beyond the active edge sensors, an additional ingredient needed to design four side buttable modules 
is the possibility of moving the wire bonding area from the chip surface facing the sensor to the backside, 
avoiding the implementation of the cantilever extruding beyond the sensor area. The feasibility of this process 
is under investigation with the FE-I3 SLID modules, where Inter Chip Vias are etched, 
employing an EMFT technology, with a cross section of 3 \mum \,x10 \mum, at the positions of the original wire bonding pads.
\end{abstract}


\end{frontmatter}


\section{Introduction}
\label{Introduction}
Different pixel modules concepts are being developed at the moment
in view of the tracker upgrades of the LHC experiments for the Phase II
of the High-Luminosity LHC (HL-LHC), in order to cope with the 
high level of occupancy and radiation damage expected. 
Detectors with minimal material budget, a larger active area
fraction and higher granularity are required, in addition to radiation hardness
up to a fluence of  $2\cdot 10^{16} \neqcm$, as for example in the case of the inner layer  
of the ATLAS pixel system. In this context thin planar pixel sensors with reduced 
inactive edges are good candidates to instruments the new trackers,
thanks to their reliability, cost-effectiveness and good charge 
collection properties at high fluences. An R\&D towards the development 
of thin n-in-p pixel modules of different types, partly including sensors with active edges,
is presented. They are designed to achieve, in conjuction
with 3D vertical integration technologies,  four side buttable
modules using Inter Chip Vias (ICV), to move the wire bonding 
area to the chip backside from the present location on the front side facing
the sensor. 
The n-in-p sensor technology is employed in all the productions described. In addition 
to an easier fabrication, requiring only a single-sided process in comparison to the
double-sided n-in-n sensors, it has also 
 been demonstrated that it achieves the same performance
before and after irradiation as the standard n-in-n technology \cite{NinPpaper,elba,philipp_pro}.
The homogeneous backside of the n-in-p sensors is also less problematic for the attachment
of a handle wafer that is an usual  step in the manifacturing of very thin pixel sensors.
\section{Thin pixel sensor productions}
\subsection{MPI HLL thin pixel productions}
To investigate the properties of thin sensors and to 
explore their potential for high energy physics applications
two successive pixel productions were carried out at MPI HLL,
employing a thinning procedure developed at this facility.
This technology makes use of a handle wafer to offer mechanical support
during the thinning phase of the active wafer and its subsequent processing
The handle wafer can then be removed by using deep anisotropic wet etching
with the oxide layer connecting the two wafers acting as etch stopper \cite{thinning}.
The first production contains ATLAS FE-I3 compatible n-in-p sensors with an active thickness of 75
$\mu$m and it has been interconnected to 
the read-out chips via the Solid Liquid Interdiffusion technology (SLID) \cite{TIPP},
developed by the Fraunhofer Institute EMFT.
This pixel production was extensively described in \cite{Hiroshima,Pixel2010,Abery}
and has been now further tested up to a fluence of $\Phi=10^{16} \neqcm$, by 
measurements of charge collection efficiency with a $^{90}$Sr radioactive source. These
were performed, as all the other charge collection measurements reported in the following,
with the ATLAS USBPix system \cite{USBPix}  inside a climate chamber keeping the 
environmental temperature at $20\,^{\circ}{\rm C}$   
for not irradiated samples and at $-50\,^{\circ}{\rm C}$ for irradiated ones.
\begin{figure}[h!]
\centering
\subfigure[]{
\includegraphics[width=0.9\columnwidth]{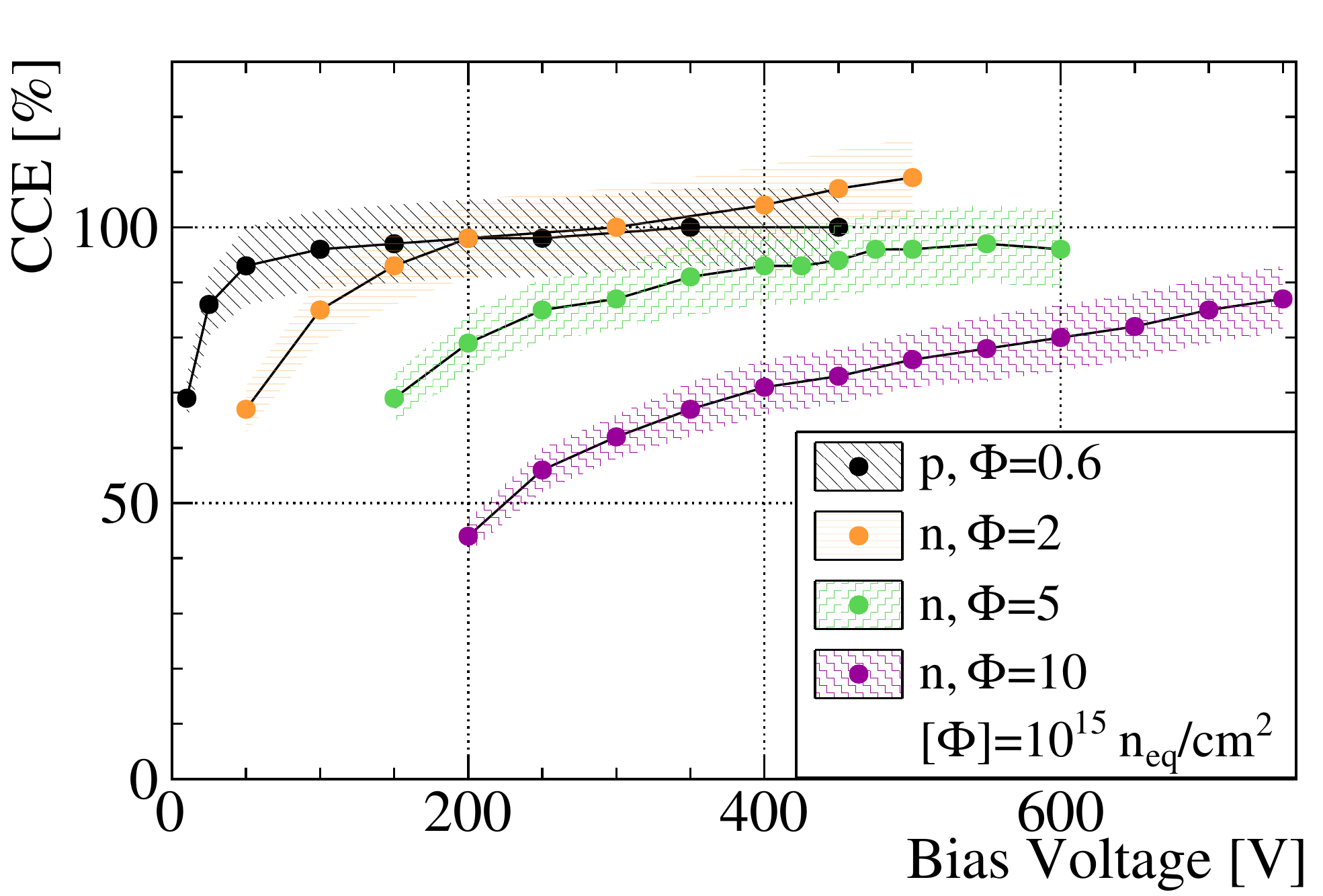}
 \vspace{-120pt}
\label{fig:SOI1CCE-a}

}

\subfigure[]{
 \hspace{0.4cm}
  \includegraphics[width=0.9\columnwidth]{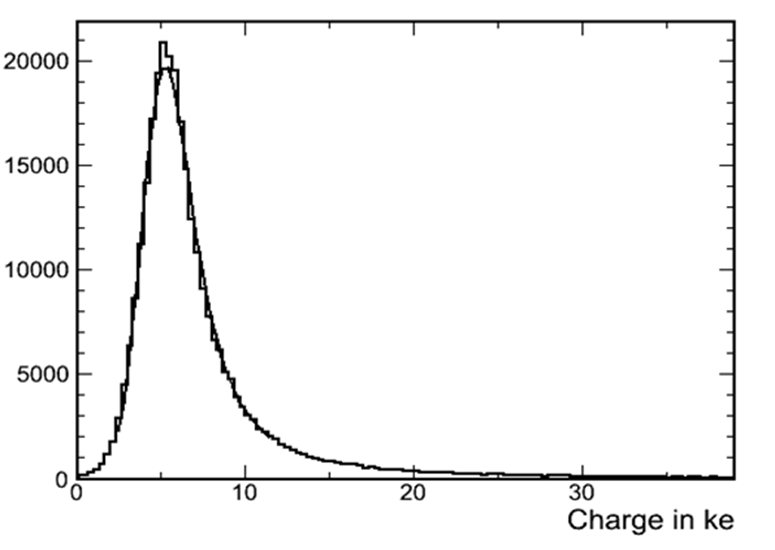}
\label{fig:SOI1CCE-b}
}
\caption{(a) Charge collection efficiency obtained with a $^{90}$Sr source for the SLID modules, 
produced with 75 \mum \, thick sensors. The collected charge is normalized to the values
obtained for the same module before irradiation. The colored bands represent
the systematic uncertainty of 5\% estimated for the normalized charge. 
(b) Landau distribution of the collected
charge for the sample irradiated at $\Phi= 5\cdot10^{15} \neqcm$, with a bias voltage of 600V.}
\label{fig:SOI1CCE}
\end{figure}

Fig.\ref{fig:SOI1CCE}(a) shows the charge collection efficiency (CCE), normalized to
the charge collected before irradiation, as a function of the bias voltage,
for different received fluences. Also at the maximum fluence of $10^{16} \neqcm$, increasing the
bias voltage up to 750V, a CCE as high as 
90$\%$ can be obtained. 
The Landau distribution of the collected charge at $\Phi= 5\cdot10^{15} \neqcm$
and a bias voltage of 600V is shown in Fig.\ref{fig:SOI1CCE}(b). The absolute value of the MPV is
of the same size as the pre-irradiation one, that lies for the different SLID modules around (4.5-5) ke, 
in agreement with the expectation for 75 \mum \, detectors \cite{bichsel}. Even if the charge collection properties 
of these very thin sensors deteriorate much less than for thick sensors after irradiation, 
the signal over threshold ratio for the FE-I3 chip  
is not high enough to operate them with full hit efficiency, given the
minimum treshold range around 2500 e achievable with the present ATLAS read-out chip.
The ATLAS FE-I4 chip \cite{FE-I4}, developed for the IBL project,
offers instead the possibility to work at thresholds as low as 1000 e,
paving the way to the use thin pixel sensors in the upgraded trackers for Phase II 
at HL-LHC. The irradiated SLID modules did not present any sign of deterioration
of the interconnection, as for example an increase in the number of disconnected channels
or of the threshold noise. These findings, even if based on a small statistics
of samples, indicate that the SLID interconnection 
is  radiation resistant up to $\Phi= 10^{16} \neqcm$.

\begin{figure}[h!]
\centering
\includegraphics[width=7cm]{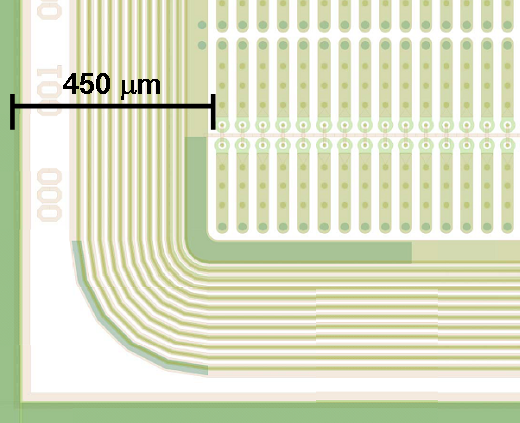}
\caption{Design of the edge region of the SOI2 pixel sensors, with twelve guard rings 
and a total distance of d$_{edge}$=450 $\mu$m from the last pixel to the dicing line.}
\label{fig:SOI2_corner}
\end{figure}

FE-I4 compatible n-in-p sensors, with an active thickness of 150 $\mu$m were obtained
with the second production at MPI HLL (indicated as SOI2 in the following) 
and they were interconnected with solder bump-bonding at the Fraunhofer Institute IZM, Berlin.
The bulk material is the same as for the SLID sensors, FZ p-type, with a resistivity of 
2 K$\Omega$cm.
The sensors are characterized by a guard ring scheme containing twelve guard rings
resulting in a distance between the last pixel implantation and the sensor edge
d$_{edge}$=450 $\mu$m (Fig.\ref{fig:SOI2_corner}).

\begin{figure}[h!]
\centering
\subfigure[]{
\includegraphics[width=0.9\columnwidth]{Noise_THR800_2.png}
\label{fig:occupancy-a}
}
\subfigure[]{
  \includegraphics[width=0.9\columnwidth]{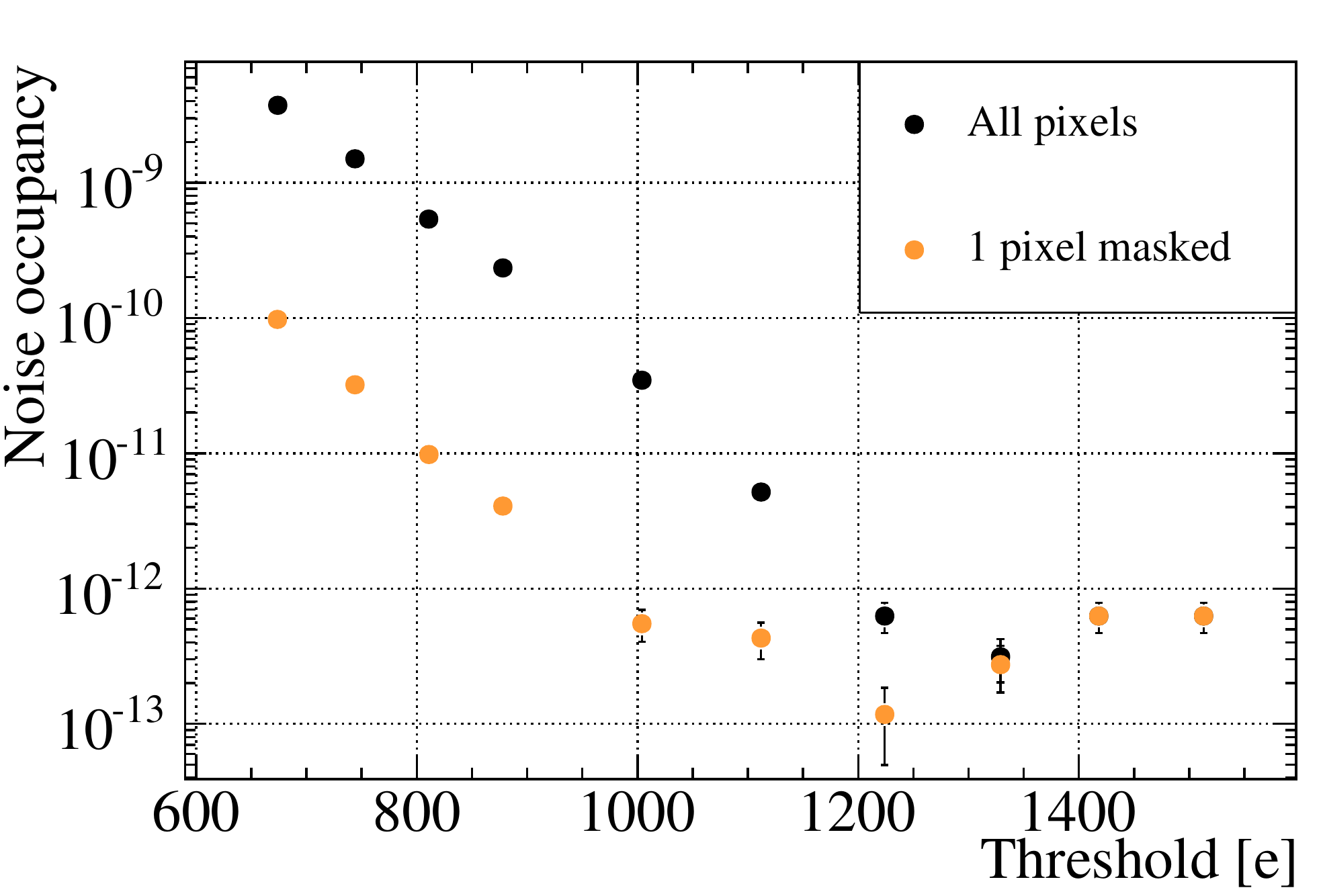}
\label{fig:occupancy-b}
}
\caption{(a) Threshold noise distribution for a not-irradiated 
SOI2 module tuned to a threshold of 800 e and a charge calibration of 
10 ToT for 15 ke. (b) Noise occupancy per pixel as
a function of the threshold.}
\label{fig:occupancy}
\end{figure}

\begin{figure}[h!]
\centering
\subfigure[]{
\includegraphics[width=0.8\columnwidth]{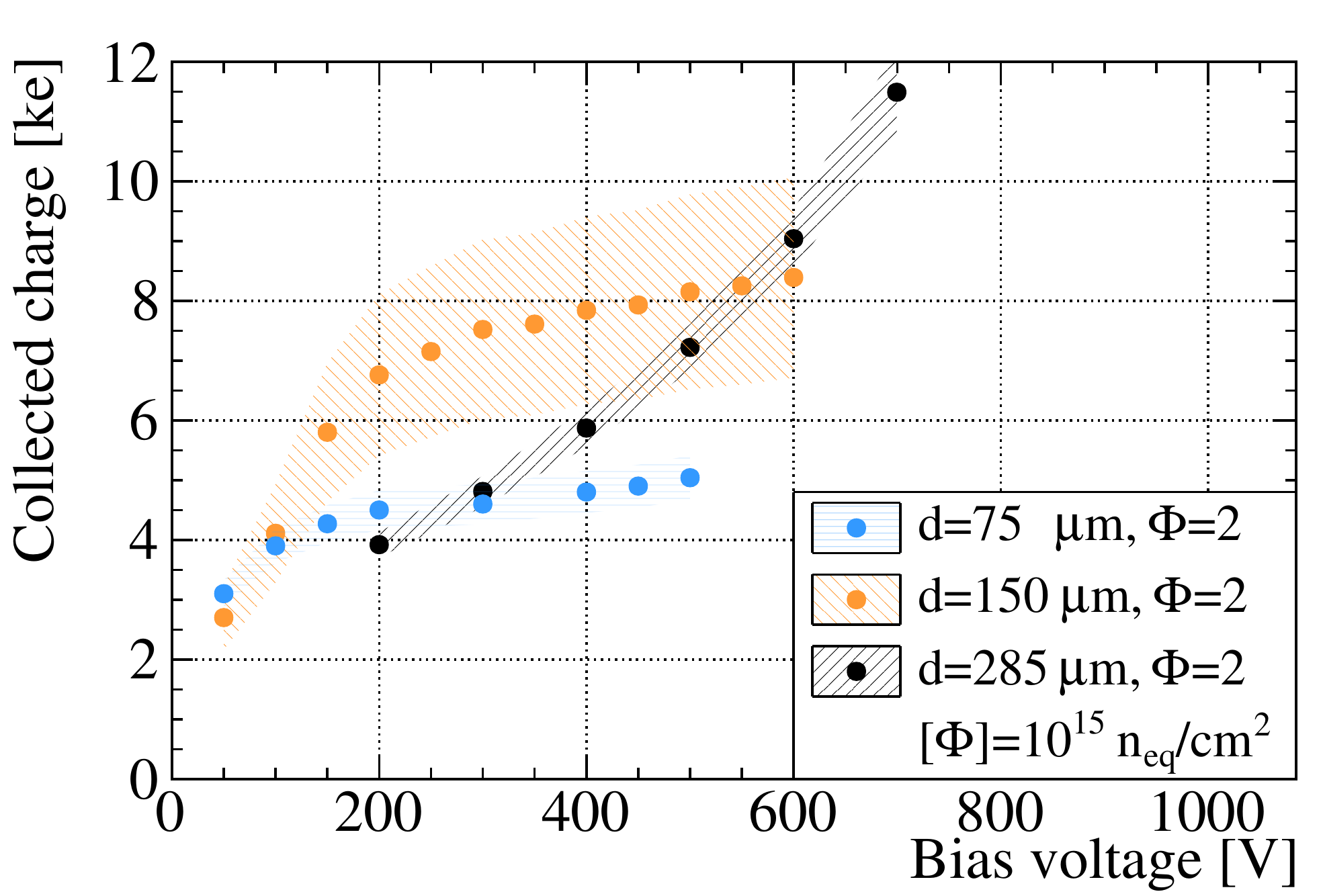}
\label{fig:chargecomparison-a}
}
\subfigure[]{
  \includegraphics[width=0.8\columnwidth]{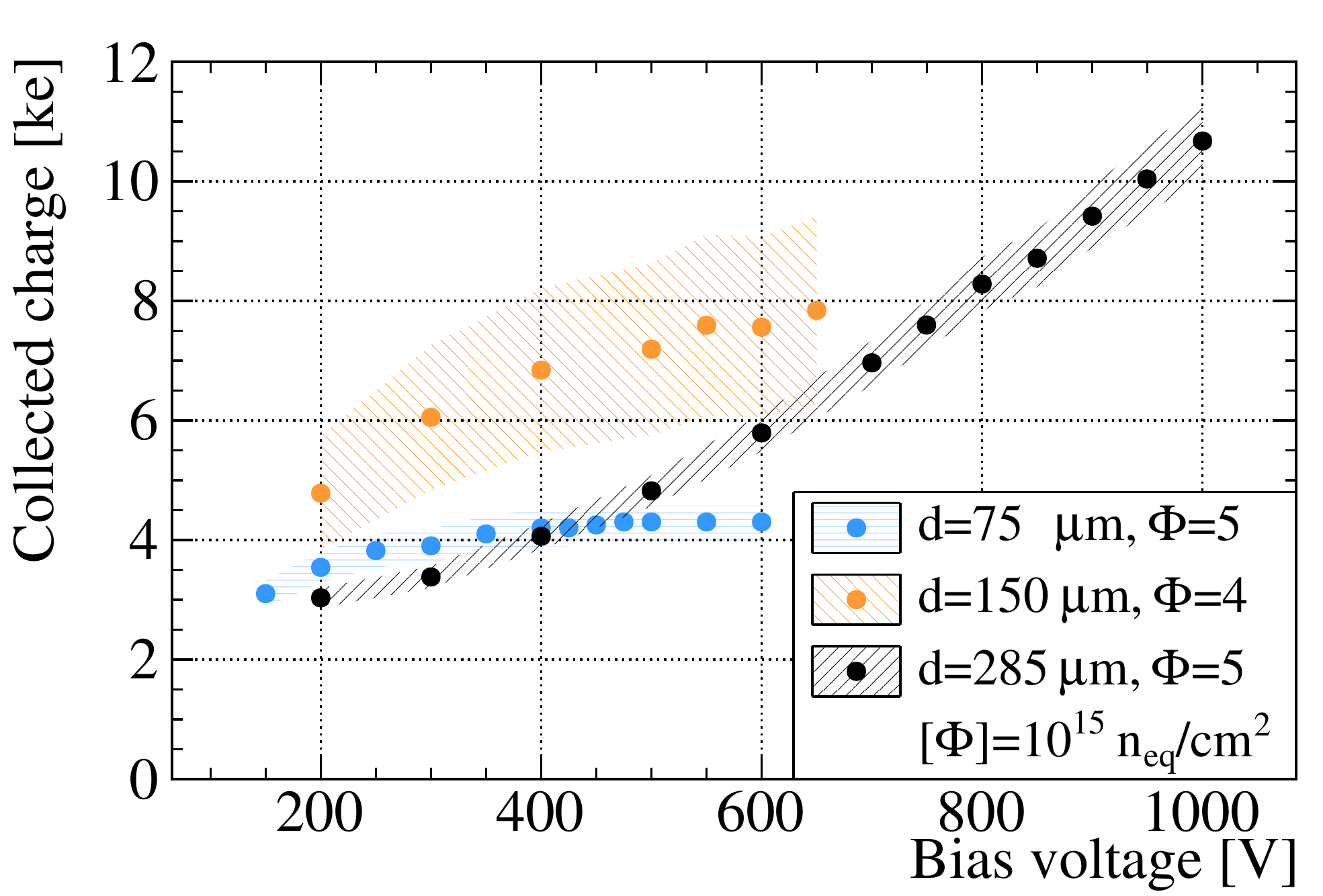}
\label{fig:chargecomparison-b}
}
\subfigure[]{
  \includegraphics[width=0.8\columnwidth]{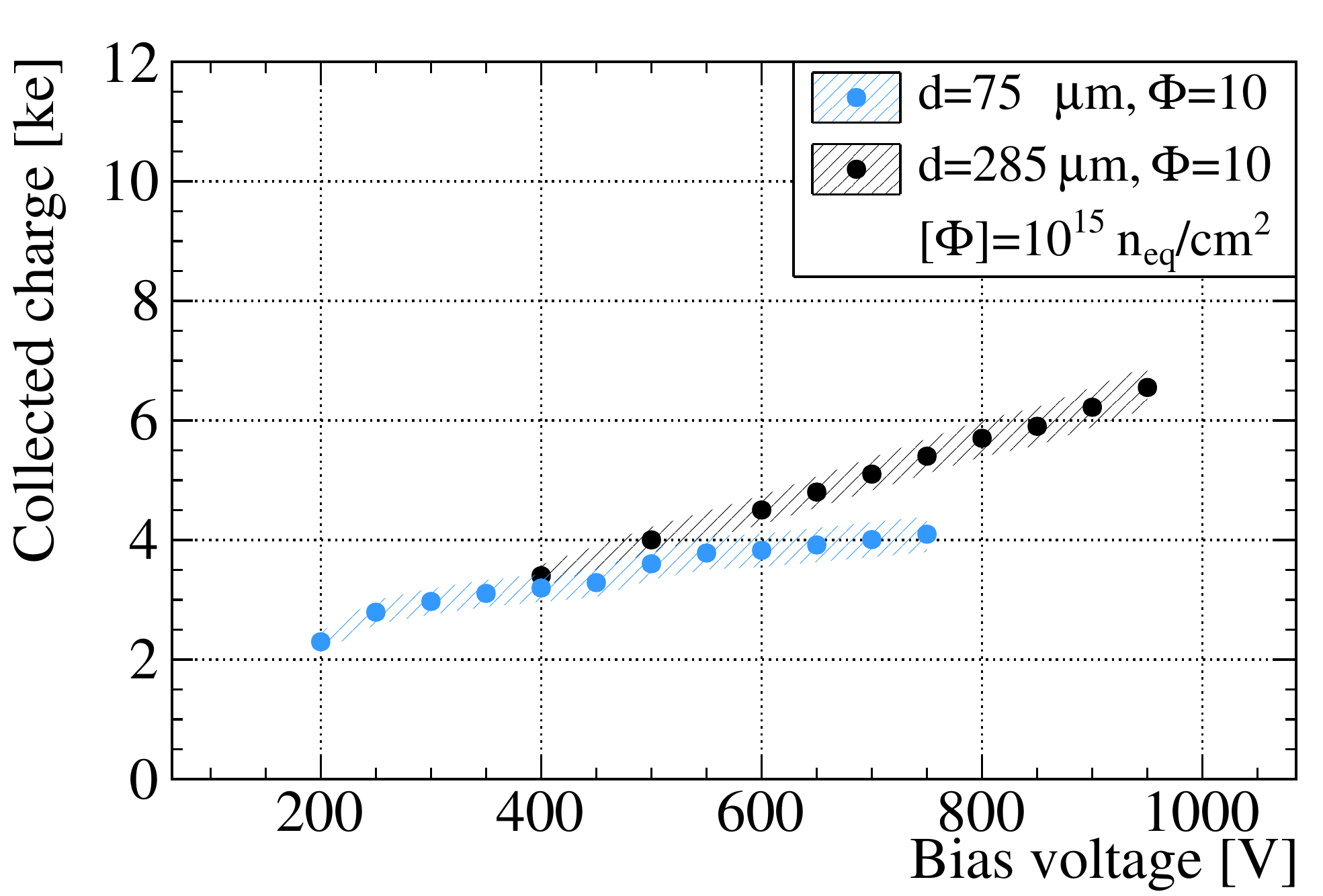}
\label{fig:chargecomparison-c}
}
\caption{Comparison of the charge collected with a $^{90}$Sr source for irradiated pixel detectors at different
thickness and received fluences: (a) results relative to $\Phi= 2\cdot 10^{15} \neqcm$, 
(b) $\Phi=(4-5)\cdot 10^{15} \neqcm$ and (c) $ \Phi=10^{16} \neqcm$. The colored bands
are representetive of the systematic uncertainties 
associated to the measurements: 7.3\%, 5.2 \%, 20\%  for the SLID sensors, 
285 $\mu$m thick sensors and the SOI2 sensors respectively. The larger 
systematic error estimated for the SOI2 
modules is due to the lack of a reliable charge calibration
for the FE-I4A chips.}
\label{fig:chargecomparison}
\end{figure}

For the FE-I4A chip no reliable ToT to charge calibration is 
available \cite{IBLmodules}, so the charge collection properties of the SOI2 assemblies are mainly 
based on ToT values, in units of the 25 ns bunch crossing clock, 
which around the tuning point are linearly proportional
to the charge. 
\begin{figure}[h!]
\centering
\includegraphics[width=\columnwidth]{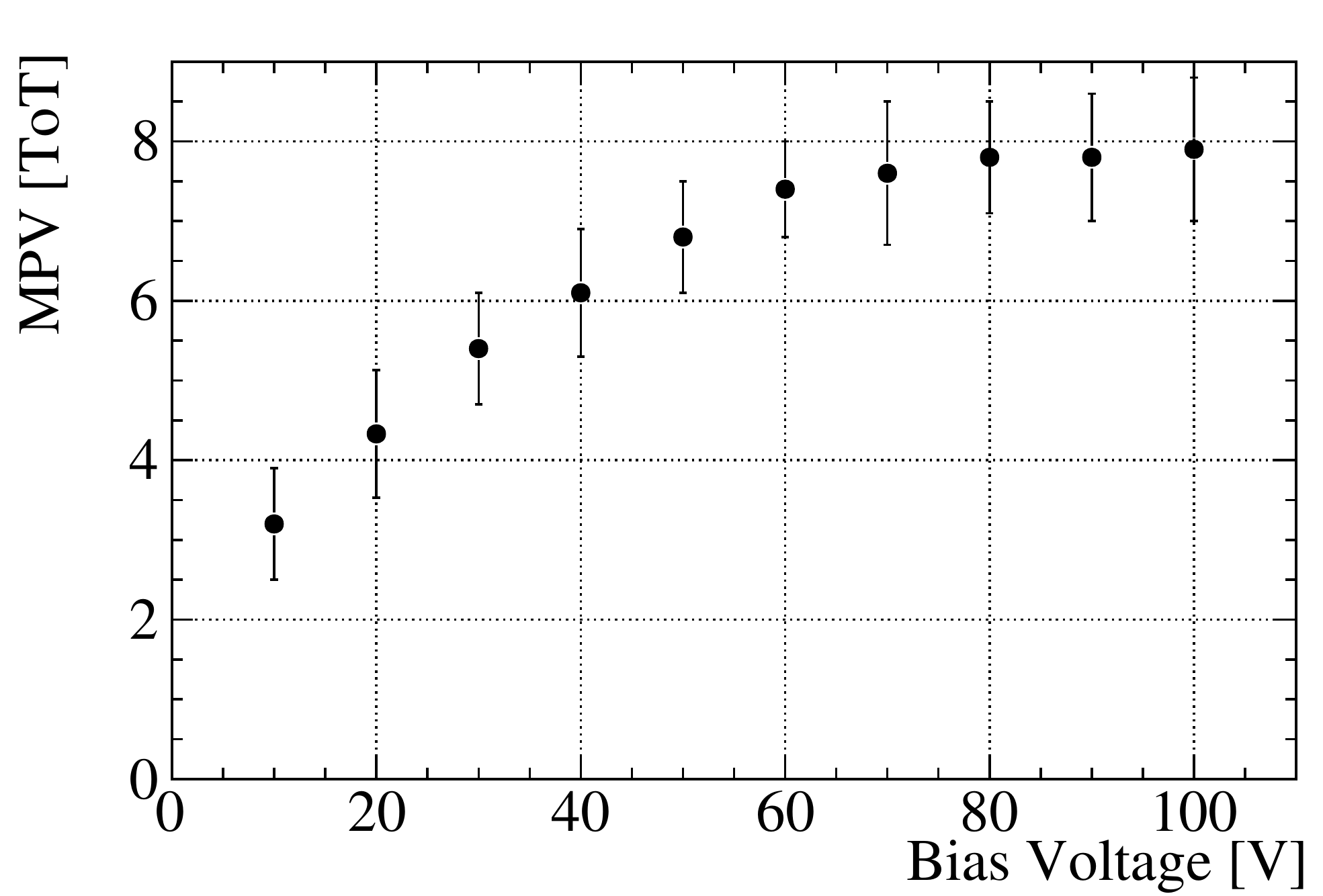}
\caption{MPV of collected charge, expressed in ToT units, 
 for a not irradiated SOI2 module, obtained from $^{90}$Sr source measurements, 
as a function of the bias voltage.}
\label{fig:MPV_Vbias}
\end{figure}
Figure \ref{fig:MPV_Vbias} shows the evolution of the most probable value (MPV) of the
measured ToT with the measured bias voltage for a not irradiated SOI2 assembly,
obtained during scans with a $^{90}$Sr source.
Given the tuning of 10 ToT for 15 ke, the saturation value of the charge corresponds to 
(12$\pm$ 2.4) ke, in agreement with the expectation for a 150 $\mu$m
thick sensor \cite{bichsel}, where a 20\% systematic error has been assigned for the FE-I4A 
calibration uncertainty.
The noise of these FE-I4 assemblies before irradiation is 
around 120 e, even when operating the chips at 
thresholds down to 800-1000 e (Fig. \ref{fig:occupancy}a).
The resulting noise occupancy is very low, as shown as a function
of the threshold in Fig. \ref{fig:occupancy}b.
A subsample of the SOI2 modules has undergone an irradiation programm, 
at a fluence of $2\cdot 10^{15}\neqcm$ with 23 MeV protons at the Karlsruhe Institute of 
Technology (KIT) and at a fluence of $4\cdot 10^{15} \neqcm$ with 800 MeV protons at the Los Alamos 
Neutrons Science Center (LANSCE).
The collected charge in scans with a $^{90}$Sr source for the SOI2 devices after irradiation
is compared  in Fig.\ref{fig:chargecomparison} to the values obtained for the 75 $\mu$m thick sensors and for 
n-in-p sensors of standard thickness, 285 $\mu$m, used as reference. 
These devices were produced
at CiS and interconnected to FE-I3 chips via bump-bonding. 
Details about their characterization before and after irradiations are reported in
\cite{NinPpaper,philipp_pro}. The comparison shows that, at least up to a fluence of 
$(4-5)\cdot 10^{15}$, where data for SOI2 are available, a higher charge may be obtained 
with these devices of intermediate thickness. Anyhow with increasing fluence 
the collected charge for different thicknesses tend to equalize, due to the effect 
of charge trapping, and at $\Phi= 10^{16} \neqcm$ (Fig.\ref{fig:chargecomparison-c}) the  75 $\mu$m  and
285 $\mu$m  shows a very comparable performance up to 600V. 

\begin{figure}[ht]
\centering
\includegraphics[width=7cm]{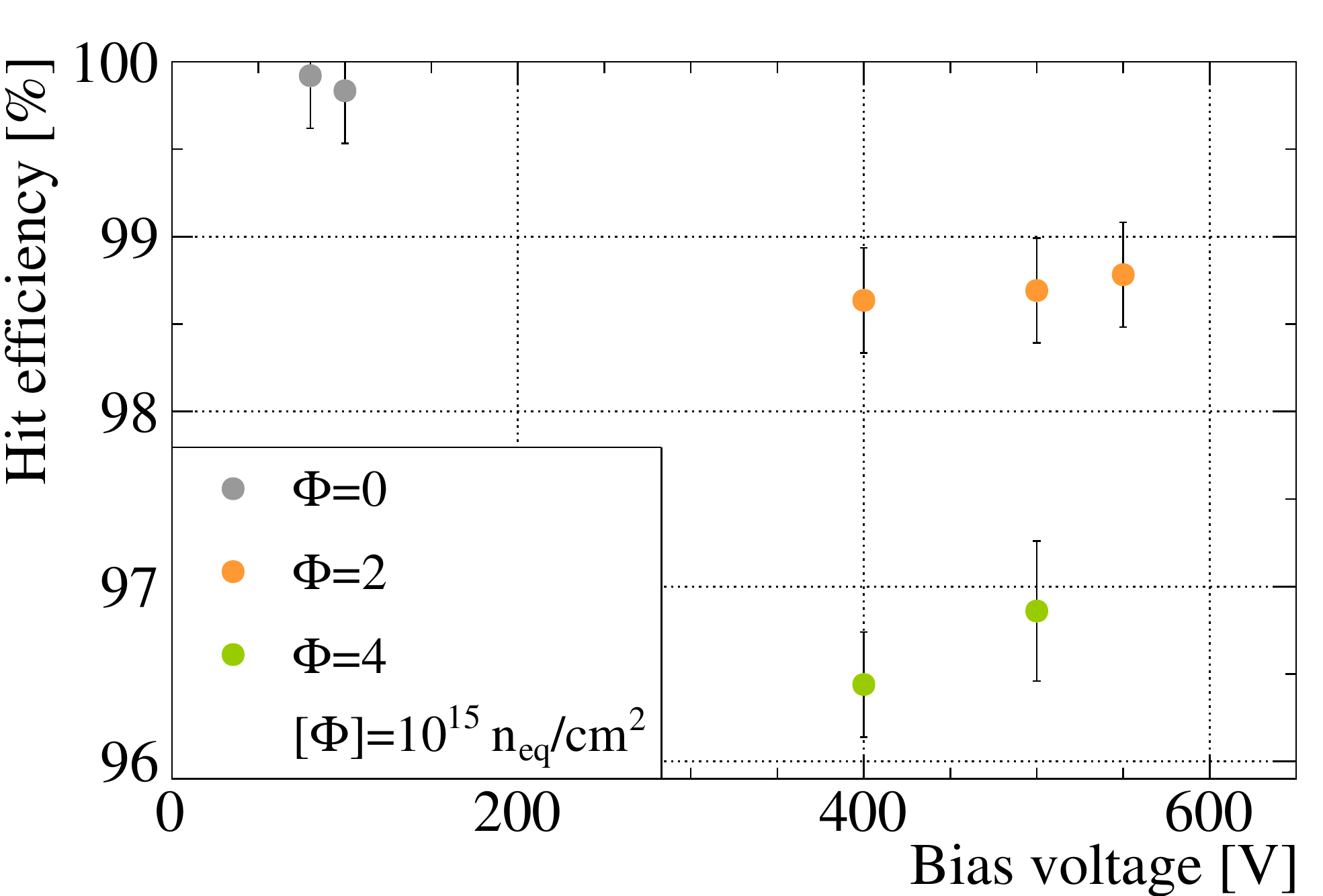}
\caption{Hit efficiency of the SOI2 modules, as a function of the applied bias voltage, for different fluences,
as obtained in beam tests with the EUDET telescope with 120 GeV pions at CERN-SPS and
5-6 GeV electrons at DESY. }
\label{fig:hitefficiency}
\end{figure}

The SOI2 modules were further studied in beam tests with 120 GeV pions at CERN-SPS and
5-6 GeV electrons at DESY, using the EUDET telescope for track reconstruction \cite{tbpaper}.
A summary of the hit efficiency as a function of the applied bias voltages, up to 
a fluence of $4\cdot 10^{15} \neqcm$ is given in Fig. \ref{fig:hitefficiency}. 

\begin{figure}[hbt]
\centering
\subfigure[]{
\includegraphics[width=0.76\columnwidth]{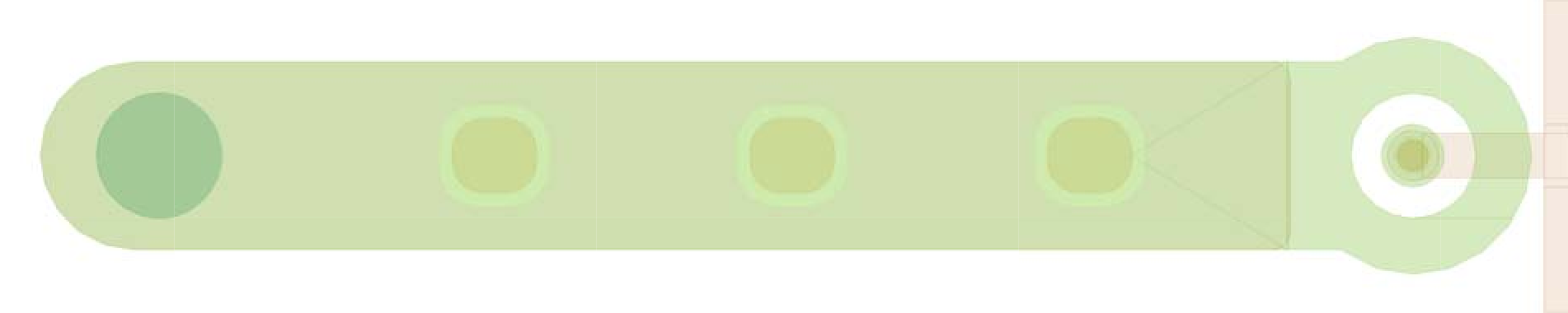}
\label{fig:occupancy-a}
}
\subfigure[]{
\includegraphics[width=0.9\columnwidth]{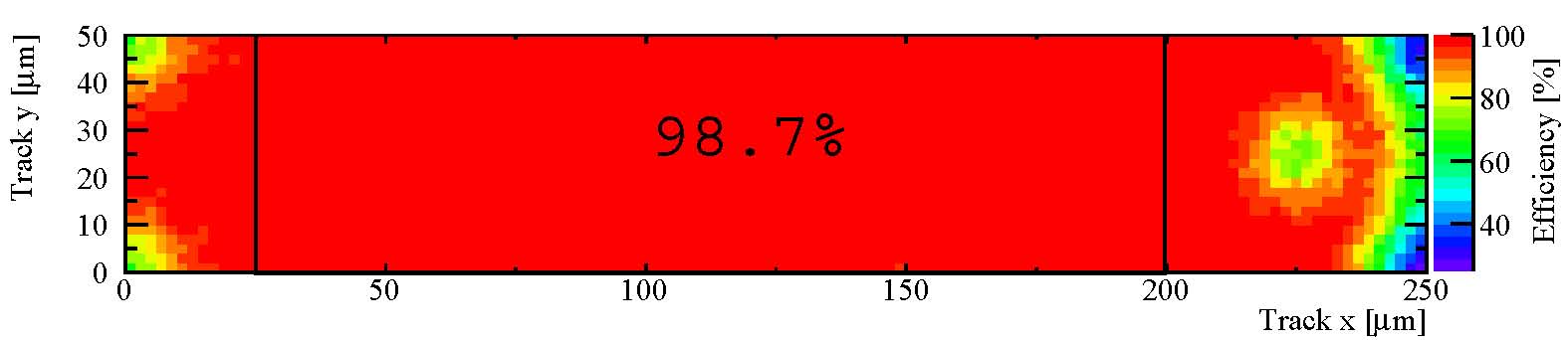}
\label{fig:occupancy-b}
}
\subfigure[]{
  \includegraphics[width=0.9\columnwidth]{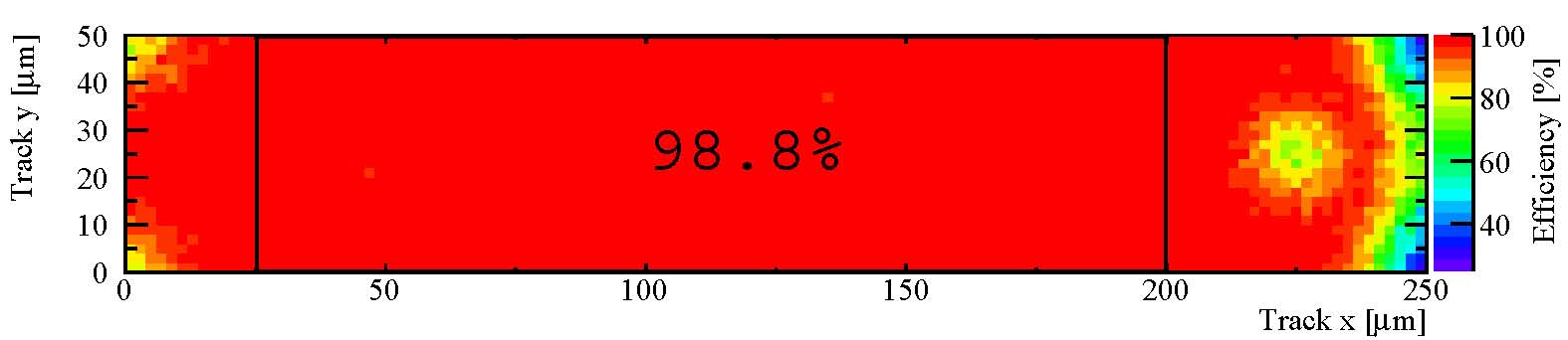}
\label{fig:occupancy-c}
}
\caption{(a) Design of the pixel cell for the SOI2 sensors, with a pitch of
50 \mum $\times$ 250 \mum.
(b) Hit efficiency map for a SOI2 module irradiated at  $\Phi= 4\cdot10^{15} \neqcm$
and biased at 400V and 500 V(c). The numbers indicated in the square 
are the hit efficiency for the central region, excluding the corners and
the bias dot.}
\label{fig:efficiencymap}
\end{figure}

Tracks extrapolated from the telescope are considered as belonging to a specific hit
if they are closer than one pixel cell pitch in the long pixel direction and three
pixel cell pitches in the short pixel cell direction. These tracks are defined
as matched. The hit efficiency is determined as the fraction of 
matched tracks to a hit in the device and it can be displayed as function of 
the position in the pixel cell, as predicted by the telescope.
Such an efficiency map is shown in Fig.\ref{fig:efficiencymap}, together with 
an image of the pixel cell geometry for the SOI2 modules. The overall efficiencies
are (96.5$\pm$0.3)$\%$ at 400V and (96.9$\pm$0.3)$\%$ at 500V, where the
errors are due to the systematic uncertainty on the track selection.    
Lower hit efficiencies are found in the corners and in the 
region corresponding to the bias rail and to the bias dot, when the beam is 
perpendicular with respect to the devices, as in this case. 
For inclined tracks part of the efficiency loss in these areas is recovered because the impinging particles
traverse also the central region of the pixel cell, where the electric field is higher.
The hit efficiencies in central region of the pixel cell, as indicated in Fig.\ref{fig:efficiencymap},
are (98.7$\pm$0.3)$\%$ at 400V and (98.8$\pm$0.3)$\%$ at 500V 

\subsection{VTT production of active edge pixel sensors}
\begin{figure}[ht]
\centering
\includegraphics[width=\columnwidth]{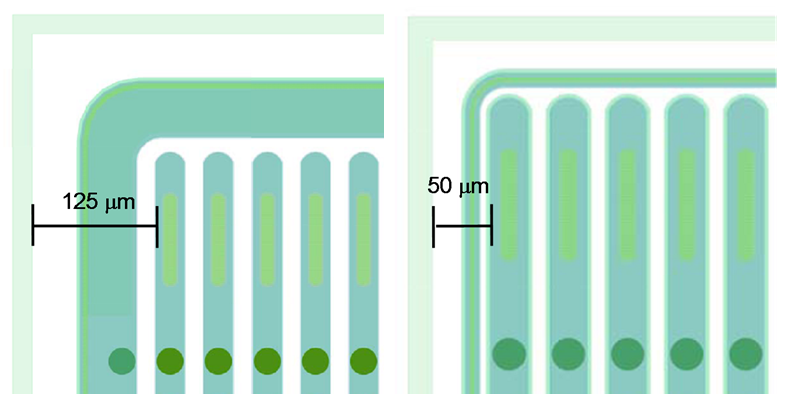}
\caption{Edge design for VTT sensors with (a) bias ring 
and guard ring with d$_{edge}$=125 \mum, 
implemented in FE-I3 and FE-I4 modules; (b) only a guard ring with d$_{edge}$=50 \mum,
implemented in FE-I3 modules.}
\label{fig:VTTedges}
\end{figure}

Active edge n-in-p pixel sensors have been fabricated at VTT 
within a multi project wafer production \cite{Kalliopuska}. 
The design of the active area of the FE-I3 and FE-I4 devices 
and the implementation of
the p-spray isolation are derived from the MPI HLL productions. 
The bulk material of the devices discussed in this paper 
is p-type FZ, with a resistivity of $\rho=10$ $k\Omega$\,cm. The sensors
were thinned to a thickness of 100 \mum. 
The fabrication of thin sensors at VTT exploits, as for the MPI HLL productions,
the use of an handle wafer as mechanical support during the grinding phase. The handle wafer 
is further needed during the etching of the trenches at the 
sensor borders, a step performed after the electrodes implantation
and prior to the final oxidation, implantation activation and
Aluminum processing \cite{VTTprocess}.  
The trenches are actived with a four-quadrant ion implantation of boron
ions, that extends the back-side junction to the vertical edges.
The interconnection of the sensors to the FE-I3 and FE-I4 chips has
been performed at VTT, with solder bump-bonding.
The results shown in the following are relative to three different implementations 
of the edge regions. The first one, meant to serve as a reference, incorporates
eleven guard rings, with the same structure as in the SOI2 production, as 
well as a bias ring, and it has d$_{edge}$=460 \mum. The second design is characterized
by only a floating guard ring and a bias ring, connected to the pixel punch-through structures,
allowing for the testability before interconnection and the grounding of each 
individual pixel, even in the case of missing bumps, after interconnection.
In this case d$_{edge}$=125 \mum \,\,(Fig.\ref{fig:VTTedges}(a)). 
The most aggressive design with active edges,
implemented only for FE-I3 sensors,
foresees one floating guard ring and d$_{edge}$=50 \mum \,\,(Fig.\ref{fig:VTTedges}(b)). Since a bias
ring structure is not present, also the bias rails and the punch-through structures
have been omitted. 

\begin{figure}[ht]
\centering
\includegraphics[width=\columnwidth]{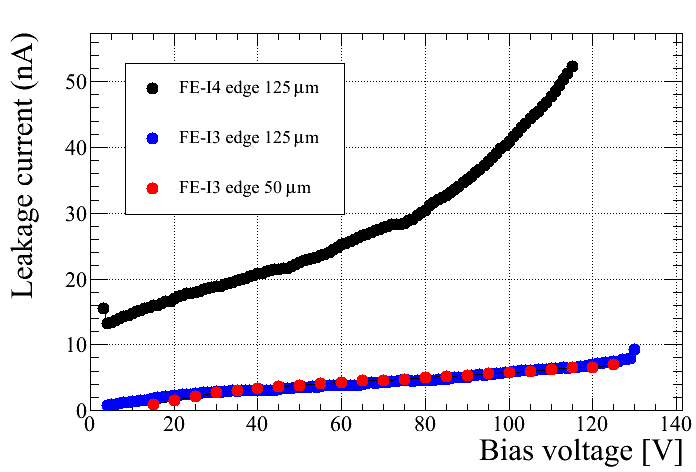}
\caption{IV curves for not irradiated FE-I3 and FE-I4 VTT modules, 
with breakdown voltages in the range 110-120V.}
\label{fig:VTTIV}
\end{figure}

Fig.\ref{fig:VTTIV} shows the IV curves of not irradiated
FE-I3 and FE-I4 modules, with breakdown voltages of the order of (110-120)V
and leakage currents well below 10 nA/cm$^2$ over all the voltage range.

\begin{figure}[h!]
\centering

\includegraphics[width=\columnwidth]{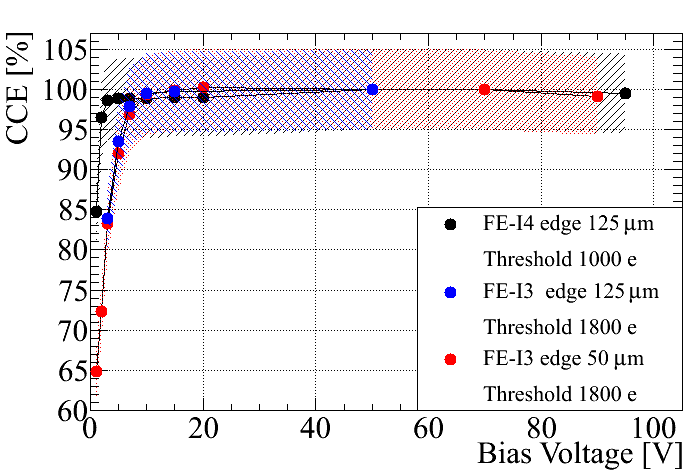}
\caption{Collected charge of VTT FE-I3 and FE-I4 modules before irradiation 
determined with a $^{90}$Sr source, normalized to the
MPV at 50V. A systematic uncertainty of 5\% is estimated for the normalized values.}
\label{fig:VTTCCE}
\end{figure}

\begin{figure}[h!]
\centering
\subfigure[]{
\includegraphics[width=\columnwidth]{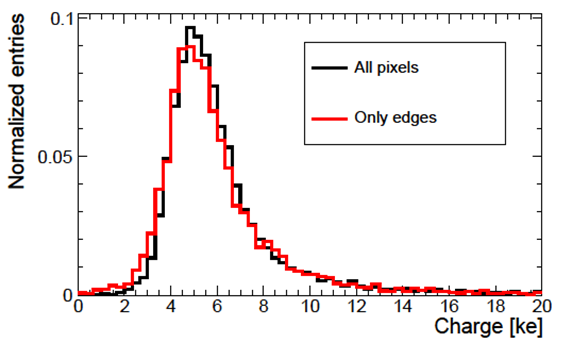}
\label{fig:vtt-a}
}
\subfigure[]{
  \includegraphics[width=\columnwidth]{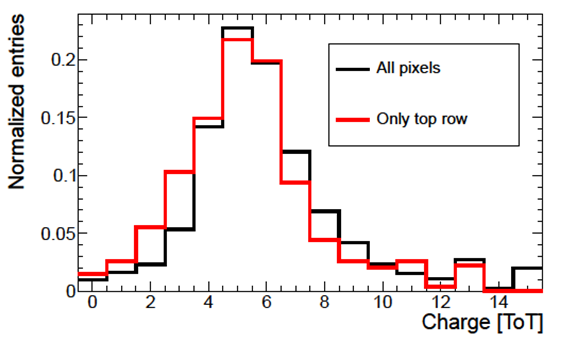}
\label{fig:vtt-b}
}
\caption{Comparison of the Landau distributions of all (black) and edge pixels (red) for the FE-I3 VTT module with d$_{edge}$=50
\mum \,(a) and for the VTT FE-I4 module with d$_{edge}$=125 \mum.  }
\label{fig:LANDAUVTT}
\end{figure}

Measurements of charge collection have been performed with these modules
after tuning of the FE-I3 samples to a threshold 
of 1800 e and of the FE-I4 sample to a threshold of 1000 e. Typical values of the 
threshold noise are measured to be around 190 e for the FE-I3 assemblies and 180 e for the FE-I4
assemblies. The MPV of the collected charge with a $^{90}$Sr source, as shown in Fig.\ref{fig:VTTCCE}, is normalized to 
the MPV at 50V, to equilibrate the different absolute scales, due to
the offset of the unknown calibration capacitance. A residual 
systematic uncertainty on the charge ratio is estimated to be 5$\%$.
The saturation of the charge collection is visible around 10V, in
agreement with the value expected from the bulk resistivity, but 
also at bias voltages as low as 5V, more of 90$\%$ of the charge
is still collected. 
A more detailed analysis of the edge pixel efficiency
has been performed, comparing the Landau distributions of these cells
to the ones obtained with the full device. The results are shown in 
Fig.\ref{fig:LANDAUVTT}, where only very small deviations are found 
between the two sets of curves, both for the FE-I3 devices with d$_{edge}$=50 \mum\,
and the FE-I4 devices with d$_{edge}$=125\mum.  The different binning in the two figures 
is due to the different ToT resolution for the two generations of chips, 8 bits for the
FE-I3 and 4 bits for the FE-I4. The performance of the edge pixels in terms of hit efficiency
will be further investigated in the near future with beam tests of these devices.

\section{Vertical Integration Technology}
\begin{figure}[ht]
\centering
\includegraphics[width=\columnwidth]{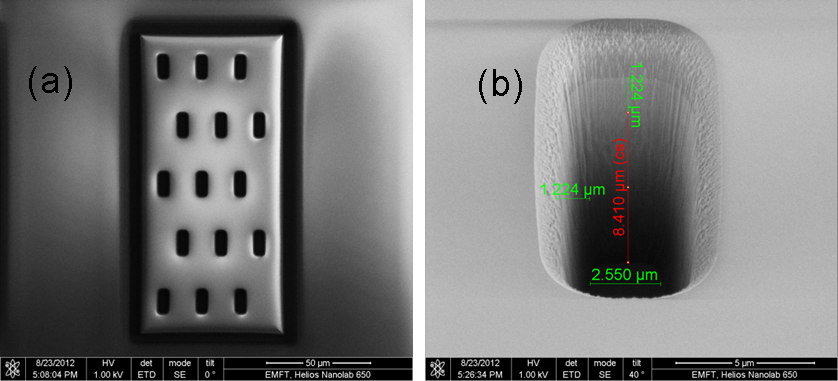}
\caption{EMFT courtesy (a) View of the wire bonding pad of the FE-I3 chip with the ICVs. Several of them are placed on the same pad 
for redundancy. A trench surrounds the ICVs to provide further isolation from the silicon bulk.
  (b) Zoom in of the ICV after the etching of the dieletric layers beneath the top aluminum layer. In red the 
depth of the etching is indicated, while in green the lateral dimensions are given. 2.550 \mum \, is the width
of the ICV at the bottom of the stack of the first dielectric layers.}
\label{fig:ICV}
\end{figure}

A second step in the R\&D activity with the modules interconnected
with SLID is the etching of Inter Chip Vias in the FE-I3 chips 
with the Via Last approach, performed at the Fraunhofer Institute EMFT,
to allow for the extraction of 
signal and services across the chip to the backside. 
In a 3D compliant design of the pixel electronics, 
Inter Chip Vias could eventually avoid the need for the
cantilever area where the wire bonding pads are presently located.
The vias cross-section has been optmized with etching trials in a FE-I3
wafer to be 3x10 $\mu$m$^2$, and the initial depth to 60 \mum.
After the isolation of the vias with Chemical Vapour Deposition (CVD) 
of silicon dioxide and the  metalization  with tungsten filling, 
the chip wafer front side is passivated and bonded to a handle wafer.
The FE-I3 wafer has then to be thinned to 50 \mum \, to expose vias and
finally new wire bonding pads are applied on the backside, connected
to the ICVs with a redistribution layer on the backside. 
The via preparation has started on the FE-I3 wafer with  
the etching of the dielectrics below the aluminum layer of the 
original wire bonding pads (Fig.\ref{fig:ICV}).
\section{Conclusions}
\label{sec:conclusions}
The characterization of pixel modules composed with thin n-in-p sensors, with a thickness range from 
75 \mum\, to 150 \mum,  interconnected to the ATLAS FE-I3 and
FE-I4 read-out chips has been presented.
The charge collected after irradiation is compared for different fluences to that obtained with
n-in-p pixel assemblies of standard thickness. In the range of fluence from 2 to $4\cdot 10^{15} \neqcm$
higher charge may be obtained with the SOI2 devices of 150 \mum\,
thickness. Anyhow, with increasing fluence, the collected
charge for different thicknesses tend to equalize, due to the effect
of charge trapping, and at $\Phi= 10^{16} \neqcm$  the
75 \mum\, and 285 \mum\ thick sensors show a very comparable performance up to
a bias voltage of 600V. Active edge pixel sensors were fabricated at VTT and
they have been studied before irradiation with a $^{90}$Sr source to compare
the behaviour of the edge and the internal pixels. Negligible differences
have been observed in terms of charge collection but
further analysis with beam tests are needed to 
determine the tracking performance of these devices and the effective
inactive edge that can be reached with them.
Finally, to otbain four side buttable modules, vertical integration
technologies are investigated. In particular Inter Chip Vias
are being prepared at the moment on FE-I3 chips, to  
allow for the extraction of signal and services across the chip to the backside.   
In a 3D compliant chip design Inter Chip Vias could eventually avoid the need for the
cantilever area, where the wire bonding pads are presently located, and this 
would lead, in conjunction with active edge sensors, to pixel modules
with no or very reduced inactive edges.

\section{Acknowledgements}
\label{sec:acknowledgment}
This work has been partially performed in the framework of the CERN RD50 Collaboration. 
The authors thank  V.~Cindro  for the irradiation at JSI,
A. Dierlamm for the irradiation at KIT 
 and S. Seidel (University of New Mexico) for the irradiations at LANSCE.
The irradiations at KIT were supported by
the Initiative and Networking Fund of the Helmholtz Association, contract HA-101 (”Physics at the Terascale”).
The irradiation at  JSI and the beam-tests have received funding from the 
European Commission under the FP7 Research Infrastructures project AIDA, grant agreement no. 262025. 
\bibliographystyle{model1-num-names}
\bibliography{<your-bib-database>}



\end{document}